\documentclass[preprint,aps,showpacs,nofootinbib,preprintnumbers,amsmath,amssymb]{revtex4-1}
\usepackage{}
\usepackage{epsfig}
\usepackage{subfigure}
\usepackage{dcolumn}
\usepackage{bm}
\usepackage{slashed}

\usepackage{graphicx,color}

\begin{document}
\title{Flavor $SU(3)$ properties of beauty tetraquark states\\ with three different light quarks}

\author{Xiao-Gang He$^{1,2,3,4}$ and Pyungwon Ko$^{4}$}
\affiliation{
$^1$Department of Physics and Astronomy, Shanghai Jiao Tong University, Shanghai 200240, 
China\\
$^{2}$Physics Division, National Center for Theoretical Sciences, Hsinchu 300 Taiwan\\
$^3$Department of Physics, National Taiwan University, Taipei 107, Taiwan\\
$^4$School of Physics, Korea Institute for Advanced Study, Seoul 02455, Korea}

\begin{abstract}
Beauty tetraquark states $X(\bar b q'q'' \bar q )$ composed of $ \bar b s u \bar d$, 
$\bar b  d s \bar u$, and $\bar b u d \bar s$, are unique that 
all the four valence quarks are different. Although the claim of existence of the first two states 
by D0 was not confirmed by data from LHCb, the possibility of such states still generated a lot of 
interests and should be pursued further.   Non-observation of $X(\bar b q'q'' \bar q )$ states by 
LHCb may be just due to a still lower production  rate than the limit of LHCb or at some 
different mass ranges. 
In this work we use light quark $SU(3)$ flavor symmetry as guideline to classify  symmetry 
properties of beauty tetraquark states. The multiplets which contain states with three different 
light quarks must be one of ${\bf \bar 6}$ or ${\bf 15}$ of $SU(3)$ representations. We study 
possible decays of such a tetraquark state into a $B$ meson and a light pesudoscalar octet meson 
by constructing a leading order chiral Lagrangian, and also provide search strategies to 
determine whether a given tetraquark state of this type belongs to ${\bf \bar 6}$ or ${\bf 15}$. 
If $X(\bar b q'q''\bar q )$  belongs to ${\bf 15}$, there are new doubly charged tetraquark states 
$\bar b u u \bar d$ and $\bar b u u \bar s$.
\end{abstract}
\maketitle

\section{introduction}

D0 collaboration recently claimed to have discovered a new tetraquark state $X(5568)$ with a mass of 
5567.8 MeV by studying the invariant energy spectrum of $B^0_s$ and $\pi^\pm$ based on $10.4$ fb$^{-1}$ 
$p \bar p$ collision at $\sqrt{s} = 1.96$ GeV\cite{d0-paper}. The significance is 5.1 $\sigma$.  
The states may be tetraquark states composed of $\bar b s u \bar d$ and $\bar b d s \bar u$. 
If true, this is the first observation of tetraquark states with four different valence quark flavors. 
This claim generated a lot of interesting theoretical studies. 
However, a few weeks later, the LHCb collaboration announced their searching results based 
on 3fb$^{-1}$ of pp collision data at $\sqrt{s} $= 7 and 8 TeV which did not find the same states\cite{lhcb}.   The existence of $X(5568)$ has not be confirmed.   In recent years, there have 
been a lot of researches in exotic states possibly composed  of four quarks, such as the 
$X$, $Y$ and $Z$ states\cite{olsen}.  There may be more different types 
of exotic states in Nature.   The $X(\bar b q'q'' \bar q )$ state, if it exist, may be a tightly 
bound state of di-quark anti-quark pair such as $[su][\bar b \bar d]$ or $[sd][\bar b \bar u]$, 
or  a ``molecular state'' of the loosely bound $B_{u,d}$ and $K$   or $B_s$ and $\pi$ mesons. 
We will refer such states generically as $X_b$ states.
Whether they are really bound states or due to some other effect should be carefully analyzed 
once data become available\cite{scattering}. Several theoretical papers have appeared discussing 
the D0 data focusing on the mass estimate, width and $J^{CP}$ nature\cite{new-papers}, 
and also possible partner states in tetraquark model\cite{partner}.  

Although the claim of existence of beauty tetraquark $X_b$ states by D0 was quickly ruled out by data 
from LHCb, these states are still interesting to study. Non-observation of these states by LHC 
may just be due to a still lower production rate than the limit of LHCb or at some different 
mass ranges.
In this work we use light quark $SU(3)$ flavor symmetry as the guideline to classify their 
symmetry properties. From the viewpoint of diquark picture of four quark states\cite{diquark},  
 they can form $0^{++}$ or $1^{++}$  states depending on whether the two anti-quarks  are 
 in flavor  symmetric or antisymmetric anti-diquark system.    Their $J^{CP}$ properties can be 
 probed by studying their decays,  such as  $X_b \rightarrow B^0_s \pi^\pm$ 
 carried out at $D0$. The $S$-wave decays of these final states will indicate the 
$\bar b su \bar d$ or $\bar b d s\bar u$ are $J^{CP} = 0^{++}$ states.
The photon from $B^* \to B \gamma$  is soft and thus not discovered.  
Then the results for $X_b$ being a $0^{++}$  or $1^{++}$ will be the same if only $B^0_s \pi^\pm$ 
measurement is concerned.  Therefore we will concentrate the properties of $J^{CP} = 0^{++}$ states 
in this paper.

The multiplets which contain states with three different light quarks must be one of the ${\bf \bar 6}$ 
or the ${\bf 15}$ of $SU(3)$ representation. 
We will study possible decays of such a tetraquark state into a B meson and a light pesudoscalar octet meson by constructing a leading order chiral Lagrangian, and also provide searching 
strategies to determine whether a given tetraquark state of this type belongs to ${\bf \bar 6}$ or ${\bf 15}$. 
If $X_b$  belongs to ${\bf 15}$, there are new doubly charged tetraquark states $\bar b u u \bar d$ and $\bar b u u \bar s$. Detection of these states will provide unique signatures for beauty tetraquark states belong to the ${\bf 15}$ representation.
We also provide other methods which can provide crucial information about whether beauty tetraquark states belong to ${\bf \bar 6}$ or
${\bf 15}$ multiplet.

\section{$SU(3)$ multiplets containing four different quarks}

The beauty tetraquark $X(\bar b q'q''\bar q)$ states are constructed from an anti-b quark $\bar b$, two light quarks $q_i q_j$ 
and one light 
anit-quark $\bar q_k$. Under the flavor $SU(3)$ symmetry, the anti-b quark is a singlet, 
the light anti-quark is  a ${\bf \bar 3}$,  and the light quark is a ${\bf 3}$. As far as the $SU(3)$ 
group structure is concerned, the irreducible states are 
\begin{eqnarray}
{\bf 3} \times {\bf 3} \times {\bf \bar 3} = {\bf 3 } + {\bf 3} + {\bf \bar 6} + {\bf 15} \;.
\end{eqnarray}

The ${\bf 3}$ representations do not contain a component with three different light quarks (anti-quark).  
Therefore the tetraquark states containing four different quarks cannot be in the ${\bf 3}$ representation. 
In the ${\bf \bar 6}$ representation, the two light quarks are 
anti-symmetric, while in the ${\bf 15}$ representation the two light quarks are symmetric. Both representations 
have states with three different quark flavors (including an anit-quark).   
Let us denote the states with four 
different quarks in ${\bf \bar 6}$ and ${\bf 15}$ as $X^{'}$  and $X$, respectively:  
\begin{eqnarray}
{\bf \bar 6}\;: &&\;\;X^{'}_{ds\bar u} \sim \bar b\bar u (ds - sd)\;,\;\;X^{'}_{su\bar d}\sim \bar b 
\bar d (s u - u s)\;,\;\;X^{'}_{ud\bar s} \sim \bar b \bar s (u d - d u)\;,\nonumber\\
{\bf 15}: &&\;\; X_{ds\bar u}\sim \bar b \bar u (d s + s d)\;,\;\;X_{su\bar d} \sim \bar b 
\bar d (s u + u s)\;,\;\;X_{ud\bar s} \sim \bar b \bar s (u d + d u)\;.
\end{eqnarray}

There are also other states come along with those with four different quarks.
The representations ${\bf \bar 6}$ and ${\bf 15}$ can be written  in tensor notations as, 
$X^k_{[i,j]}$ and  $ X^k_{\{i,j\}}$, respectively.  Here $[i,j]$ and $\{i,j\}$ indicate  anti-symmetric 
and symmetric combinations under exchange of flavor indices $i$ and $j$.   Both $X({\bf \bar 6})$ and 
$ X({\bf 15})$ are traceless, that is,   $X^i_{[i,j]} = X^i_{\{i,j\}} = 0$. Writing $X^k_{[i,j]}$ 
in terms of the properly normalized component fields,   
the component fields in ${\bf \bar 6}$ can be written as
\begin{eqnarray}
&&X^1_{[2,3]} = {1\over \sqrt{2}} X^{'}_{ds\bar u}\;,\;\;X^2_{[3,1]}= {1\over \sqrt{2}}X^{'}_{su\bar d}\;,\;\;X^3_{[1,2]} = {1\over \sqrt{2}}X^{'}_{ud\bar s}\;,\nonumber\\
&&X^1_{[1,2]} = X^3_{[2,3]} = {1\over 2} Y^{'}_{(u\bar u, s\bar s)d}\;,\;\;
X^1_{[3,1]} = X^2_{[2,3]} = {1\over 2} Y^{'}_{(u\bar u,d\bar d)s}\;,\;\;X^2_{[1,2]} = X^3_{[3,1]} = {1\over 2} Y^{'}_{(d\bar d, s\bar s)u}\;.
\end{eqnarray}
The component fields in ${\bf 15}$ are
\begin{eqnarray}
&&X^1_{\{2,3\}} = {1\over \sqrt{2}} X_{ds\bar u}\;,\;\;X^2_{\{3,1\}}= {1\over \sqrt{2}}X_{su\bar d}\;,\;\;X^3_{\{1,2\}} = {1\over \sqrt{2}}X_{ud\bar s}\;,\nonumber\\
&&X^1_{\{1,1\}} = ({Y_{\pi u}\over \sqrt{2}})+ {Y_{\eta u}\over \sqrt{6}})\;,\;\;
X^1_{\{1,2\}} = {1\over \sqrt{2}}({Y_{\pi d}\over \sqrt{2}})+ {Y_{\eta d}\over \sqrt{6}})\;,\;\;X^1_{\{1,3\}} = {1\over \sqrt{2}}({Y_{\pi s}\over \sqrt{2}}+ {Y_{\eta s}\over \sqrt{6}})\;,\nonumber\\
&&X^2_{\{2,1\}} = {1\over \sqrt{2}}(-{Y_{\pi u}\over \sqrt{2}}+ {Y_{\eta u}\over \sqrt{6}})\;,\;\;X^2_{\{2,2\}} = (-{Y_{\pi d} \over \sqrt{2}}+ {Y_{\eta d} \over \sqrt{6}})\;,\;\;
X^2_{\{2,3\}} = {1\over \sqrt{2}}(-{Y_{\pi s}\over \sqrt{2}}+ {Y_{\eta s}\over \sqrt{6}})\;,\nonumber\\
&&X^3_{\{3,1\}} = -{1\over \sqrt{3}}Y_{\eta u}\;,\;\;X^3_{\{3,2\}} = -{1\over \sqrt{3}}Y_{\eta d}\;,\;\;
X^3_{\{3,3\}} = -{1\over \sqrt{3}}Y_{\eta s}\;,\nonumber\\
&&X^1_{\{2,2\}} = Z_{dd\bar u}\;,\;\;X^1_{\{3,3\}} = Z_{ss\bar u }\;,\;\;X^2_{\{1,1\}} = Z_{uu \bar d}\;,\nonumber\\
&&X^2_{\{3,3\}} = Z_{ss\bar d}\;,\;\;X^3_{\{1,1\}} = Z_{uu\bar s}\;,\;\;Z^3_{\{2,2\}} = Z_{dd\bar s}\;.
\end{eqnarray}

It is clear that if $SU(3)$ flavor symmetry is a good (approximate) symmetry 
to describe the  properties of beauty tetraquark $X_b$ states, there are other states come along with the
states with four different quarks depending on which representation the tetraquark states with four different quarks belong to.
If a tetraquark state composed of four different quarks is found, 
it is important to determine whether it belongs to ${\bf \bar 6}$ or ${\bf 15}$.
It is interesting to note that 
if $X_b$  belongs to ${\bf 15}$, there are new doubly charged tetraquark states $\bar b u u \bar d$ and $\bar b u u \bar s$. Detection of these states will provide unique signatures for beauty tetraquark states belong to the ${\bf 15}$ multiplets.

\section{Chiral Effective Theory for $X(\bar b q' q''\bar q)$}

In the following we study possible decays of a tetraquark $X_b$ state into a $B$ meson and a light 
pseudoscalar octet meson by constructing a leading order chiral Lagrangian, and also provide searching 
strategies to determine whether a given tetraquark state of this type belongs to ${\bf \bar 6}$ or ${\bf 15}$. 
One can construct a chiral perturbation theory\cite{chiral-theory} to describe a $X_b$ decays into a $B$ meson 
and a light octet pseudoscalar $\Pi$.    Here $\Pi$ is the normalized  pseudoscalar octect containing 
$\pi^0,\; \pi^\pm,\; K^\pm,K^0,\; \bar K^0,\;\eta$ fields,
\begin{eqnarray}
\Pi = \left (\begin{array}{ccc}
\pi^0/\sqrt{2}+ \eta/\sqrt{6}&\pi^+&K^+\\
\pi^-&-\pi^0/\sqrt{2}+\eta/\sqrt{6}&K^0\\
K^-&\bar K^0&-2\eta/\sqrt{6}
\end{array}
\right ).
\end{eqnarray}

Under chiral $SU(3)_L\times SU(3)_R$, the Goldstone boson field $\Sigma \equiv  
{\rm exp}(2 i \pi (x) / f) $ transforms  as $\Sigma \rightarrow L \Sigma R^\dagger$ with 
$\pi(x) = \Pi/\sqrt{2}$. $f \approx 93$ MeV is the pion decay constant. 
It is convenient to define another field $\xi (x)$ by $\Sigma (x) \equiv \xi (x)^2$, 
which transforms as  $\xi(x) \rightarrow L \xi(x) U^\dagger (x) = U(x) \xi(x) R^\dagger$ under global chiral 
$SU(3)_L\times SU(3)_R$ transformation.   
The 3$\times$3 matrix field  $U (x)$ depends  on Goldstone fields $\pi(x)$ as well as 
on the global $SU(3)$ transformation  matrices  $L$ and $R$,  thereby being $x$-dependent  
field describing hidden local SU(3) transformation in chiral symmetric theories.   
The fields $X_b$ and  $B_i = ( B_u , B_d , B_s )$ transforms as $\bar{\bf 6}$ or ${\bf 15}$, 
and ${\bf 3}$ of $SU(3)$, respectively: 
\[
X_{jk}^i \to U^i_{~i'} X^{i'}_{j' k'} (U^\dagger)_{~j}^{j'} (U^\dagger)_{~k}^{k'} , 
\ \ \ B_i \rightarrow U_i^{~j} B_j . 
\]

It is convenient to define vector and axial vector fields ($V_\mu$ and $A_\mu$) with following properties 
under chiral transformations:
\begin{eqnarray}
&&V_\mu =  \frac{1}{2} \left( \xi^\dagger \partial_\mu \xi + \xi \partial_\mu \xi^\dagger \right) \rightarrow U V_\mu U^\dagger + U \partial_\mu U^\dagger \;, \;\;A_\mu =  \frac{1}{2} 
\left( \xi^\dagger \partial_\mu \xi  - \xi \partial_\mu \xi^\dagger \right)  \rightarrow U A_\mu 
U^\dagger\;.
\end{eqnarray}
Note that the vector field $V_\mu$ transforms like a gauge field for local $SU(3)$.  

For the pionic transitions $X_b \rightarrow B \pi$, the mass difference between $X_b$ and $B$ 
is very small and the pion is soft. Therefore we can use the heavy  particle effective 
theory for $X_b$ and $B$, assuming their velocity $v$ is conserved \cite{chiral-theory}.  
We should keep in mind that $X_b$ and $B$ fields have $v$-dependence so that they 
are actually $X(v)$  and $B(v)$.  The $SU(3)$ properties of the relevant fields are given by 
$X^i_{j k}$ and $B_{i}$.  One then has 
\begin{equation}
{\cal L} = \overline {B} i v \cdot D B + \overline{ X} iv\cdot DX 
 - \overline{X} M_\xi X  - \overline{B} v\cdot A X
 \label{eq:lagrangian}
\end{equation}
where we assume that all the $SU(3)$ flavor indices are contracted appropriately 
in order to respect  chiral symmetry.
\begin{eqnarray}
(D_\mu B)_i & = & \partial_\mu B_i+ (V_\mu)^{i'}_iB_{i'}\;,\nonumber\\
(D_\mu X)_{ij}^k & = & \partial_\mu X_{ij}^k +(V_\mu)_{k'}^k X_{ij}^{k'} 
- (V_\mu)_{i}^{i'}X_{i'j}^k-(V_\mu)_{j}^{j'}X_{ij'}\;.
\end{eqnarray}
Here $X_{ij}^k$ can be $X_{[i,j]}^k$ or $X_{\{i,j\}}^k$ depending on whether it is a 
${\bf \bar 6}$ or a ${\bf 15}$.
When constructing the above leading order chiral Lagrangian,  we have assumed that the Lagrangian respects parity, charge-conjugation, Lorentz symmetries, and chiral symmetry  
in terms of local $SU(3)$  hidden gauge symmetry mentioned before.  
The explicit forms of the last two operators shall be explained below in detail. 
Note that the velocity $v$ dependent scalar fields $X_b$ and $B$ we are using now 
include a wavefunction normalization factors $\sqrt{M_X}$ and $\sqrt{M_B}$ respectively. 
Therefore their mass dimensions are $[X] = [B] = 3/2$, and not $1$, in this formulation. 

We would like to comment that if $X_b$ turns out to be a multiplet with $J^{P} = 1^{++}$ the analysis is similar. 
One just needs to replace the $0^{++}$ state for $X$ by a $1^{++}$ state $X_\mu$ with a Lorentz index. 
In our discussions below we will  concentrate on the $SU(3)$ symmetry properties and 
therefore will not show the Lorentz indices which should always be contracted. 

In the exact $SU(3)$ limit, all the $X_b$ states in each representation have the same mass, 
either $m_{\bar 6}$  or $m_{15}$, which have been removed in the velocity-dependent heavy 
particle effective theory formalism. 
This degeneracy is broken by light quark masses, which is schematically represented as 
$\overline{X} M_\xi X $ in Eq. (\ref{eq:lagrangian}).  Here $M_\xi \equiv \xi M \xi + \xi^\dagger M \xi^\dagger $ 
which transforms as an $SU(3)$ octet:  $M_\xi \rightarrow U(x) M_\xi U^\dagger (x)$ and 
$M = {\rm diag} ( m_u , m_d , m_s ) \rightarrow L M R^\dagger$. 
Including all possible combinations, the corrections  come from the following terms
\begin{eqnarray}
&&\mbox{For ${\bf \bar 6}$}:\;\;\;{1\over 2}\left (a^{'} \overline{X}_k^{[i,j]}(M_\xi)_j^l X_{[i,l]}^k + b^{'} \overline{X}_k^{[i,j]}(M_\xi)^k_l X_{[i,j]}^l \right )\;,\nonumber\\
&&\mbox{For ${\bf 15}$}:\;{1\over 2} \left (a\overline{X}_k^{\{i,j\}}(M_\xi)_j^l X_{\{i,l\}}^k + b \overline{X}_k^{\{i,j\}}(M_\xi)^k_l X_{\{i,j\}}^l\right ) \;, \label{mass-s}
\end{eqnarray}
The parameters $a(a')$ and $b(b')$ have dimensionless $\sim O(1)$ couplings in this formalism.

\section{$X(\bar b q'q''\bar q)$ mass splitting and two-body decays}

Expanding the expressions in Eq.(\ref{mass-s}), we obtain the masses of the states in ${\bf \bar 6}$ and
${\bf 15}$ representations including the leading $SU(3)$ breaking corrections from a non-zero 
strange quark mass $m_s$: 
\begin{eqnarray}
&&m(X^{'}_{ds\bar{u}}) = m(X^{'}_{su\bar{d}} ) = m_{\bar{6}} + \frac{1}{2} a^{'} m_s\;,\nonumber\\
&&m(X^{'}_{ud\bar{s}}) = m_{\bar{6}} + b^{'} m_s\;,\nonumber\\
&&m(Y^{'}_{(uu,dd)\bar{s}} ) = m_{\bar{6}} +  \frac{1}{2} a^{'} m_s 
+ \frac{1}{2} b' m_s\;, \nonumber\\
&&m(Y^{'}_{(u\bar u, s\bar s)d}) = m_{\bar{6}} + {1\over 4} a^{'}m_s\;,\\
&&m(Y^{'}_{(d\bar{d},s\bar{s})u}) = m_{\bar{6}} +  \frac{1}{4} a^{'} m_s 
+ \frac{1}{2} b^{'} m_s\;,\nonumber \label{sum6}
\end{eqnarray}
for the ${\bf \bar{6}}$ representation,
and
\begin{eqnarray}
&&m(Z_{uu\bar{d}} ) = m(Z_{dd\bar{u}}) = m(Y_{\pi\bar{u}}) = m( Y_{\pi\bar{d}}) = m_{15}\;,\nonumber\\
&&m(X_{su\bar{d}}) = m ( X_{ds\bar{u}}) = m(Y_{\pi\bar{s}}) = m_{15} + \frac{1}{2} a m_s
\nonumber\;, \\
&&m(Z_{ss\bar{u}}) = m(Z_{ss\bar{d}}) = m_{15} + a m_s\;,\nonumber\\
&&m( X_{ud\bar{s}}) = m(Z_{uu\bar{s}}) = m(Z_{dd\bar{s}}) = m_{15} + b m_s\;, 
\\
&&m(Y_{\eta\bar{u}}) = m(Y_{\eta\bar{d}}) = m_{15} + \frac{1}{3} a m_s + \frac{2}{3} b m_s\;,\nonumber\\
&&m(Y_{\eta\bar{s}}) = m_{15} + \frac{5}{6} a m_s + \frac{2}{3} b m_s \;,\nonumber \label{sum15}
\end{eqnarray}
for the ${\bf 15}$ representation.  

The above equations imply that, when $u$ and $d$ quark mass contributions are neglected, 
$X^{'}_{ds\bar u} (X_{ds\bar u})$ and $X^{'}_{su\bar d}$($X_{su\bar d}$) are degenerate 
in mass. But the correction to $X^{'}_{ud\bar s}$ 
($X_{ud\bar s}$) will be different.  
$m_{\bar 6}$ and $m_{15}$ are the common masses for
the states in the ${\bf \bar 6}$ and ${\bf 15}$ representations respectively. 
The states $Z_{uu\bar d}$, $Z_{dd\bar u}$, $Y_{\pi \bar u}$ 
and $Y_{\pi \bar d}$  in the {\bf 15} representation do not receive any corrections and are 
degenerate in mass. Mass measurement for any one of them will determine $m_{15}$. While 
$m_{\bar 6}$ can be determined by measuring $2 m(Y^{'}_{(u\bar u, s\bar s)d})  - m(X^{'}_{ds\bar{u}})$.

One can derive sum rules for the masses of tetraquark states in the ${\bf \bar{6}}$ or ${\bf 15}$ analogous to Gell-Mann-Okubo formulae. 
The sum rules obtained from Eqs.(\ref{sum6}) and (\ref{sum15}) will  provide useful information 
to reconstruct the masses such as the mass of $X(X')_{ud\bar s}$. 
We will come back to this later.

The properties of $X(\bar b q'q'' \bar q)$ can be probed by studying their decays. Among the simplest dacy modes are the two-body
decay with a $B$ meson and a light pesudoscalar meson in $\Pi$ which can be obtained from the leading order chiral Lagrangian constructed in the previous section. 
The terms responsible for these decays come from the last operator of Eq.(\ref{eq:lagrangian}), which are renormalizable interactions with dimensionless couplings $\alpha$ and  
$\alpha^{'}$  of $\sim O(1)$:
\begin{eqnarray}
\alpha^{'} \overline{B}^i (v\cdot A)_k^j X_{[i,j]}^{k}\;,\;\;
\alpha  \overline{B}^i (v\cdot A)_k^j X_{\{i,j\}}^{k}\;,\;\;
\end{eqnarray}
for $X_b$ in the ${\bf \bar{6}}$ and ${\bf15}$ representations, respectively.  
Note that there is a unique operator for $X_b^{'}$ and $X_b$ 
that is relevant to the pionic transitions,  $X_b\rightarrow B \Pi$, at leading order in chiral expansion.  

The  term inducing $X_b$ to a triplet $B_i$ and the pseudoscalar octet $\Pi^i_j$ from 
the above can be parameterized as
\begin{eqnarray}
\alpha^{'} X^k_{[i,j]} \overline{B}^i \Pi^j_k\;,\;\;  \alpha X^k_{\{i,j\}} \overline{B}^i \Pi^j_k\;.
\end{eqnarray}
Depending on values for the beauty tetraquark $X_b$ mass, there are different possible decay modes 
allowed. If the mass of  a beauty tetraquark  has a mass around 5568 MeV as what D0 found, only 
$X_b \to B_i \pi$ modes are open kinematically. In this case, expanding the above we obtain the decay 
amplitudes for $A(X (X') \to B_i \Pi^j_k)$ with a $\pi$   in the final states :
\begin{eqnarray}
&&A(X^{'}_{su\bar d} \to B^0_s \pi^+ ) = - A(X^{'}_{ds\bar u} \to B^0_s \pi^-) = A(Y^{'}_{(u\bar u, d\bar d)s} \to B^0_s \pi^0) = {1\over \sqrt{2}} \alpha^{'}\;,\nonumber\\
&&A(Y^{'}_{(u\bar u, s\bar s)d} \to B^0 \pi^0) =A(Y^{'}_{(d\bar d, s\bar s)u} \to B^+ \pi^0) =  -{1\over 2\sqrt{2}} \alpha^{'}\;,\\
&&A(Y^{'}_{(u\bar u, s\bar s)d} \to B^+ \pi^-) =-A(Y^{'}_{(d\bar d, s\bar s)u} \to B^0 \pi^0) =  {1\over 2} \alpha^{'}\;,\nonumber
\label{6pi}
\end{eqnarray}
for the states in the $\bar 6$, and 
\begin{eqnarray}
&&A(X_{su\bar d} \to B^0_s \pi^+ ) = A(X_{ds\bar u} \to B^0_s \pi^-) = A(Y_{\pi s} \to B^0_s \pi^0) =  {1\over \sqrt{2}}\alpha\;,\nonumber\\
&& A(Y_{\pi d}\to B^0 \pi^0) = A(Y_{\pi u}\to B^+ \pi^0) ={1\over 2} (1+{1\over \sqrt{2}})\alpha\;,\nonumber\\
&&A(Y_{\pi d}\to B^+ \pi^-)=-A(Y_{\pi u} \to B^0\pi^+)  = {1\over 2} \alpha\;,\\
&&A(Y_{\eta u} \to B^+ \pi^0) = - A(Y_{\eta d} \to B^0 \pi^0) = {1\over 2\sqrt{3}}(1-{1\over \sqrt{2}})\alpha\;,\nonumber\\
&&A(Y_{\eta u}\to B^0 \pi^+) = A(Y_{\eta d} \to B^+\pi^-) = {1\over 2\sqrt{3}}\alpha\;,\nonumber\\
&&A(Z_{dd\bar u} \to B^0 \pi^-) = A(Z_{uu\bar d} \to B^+ \pi^+) = \alpha\;,\nonumber
\label{15pi}
\end{eqnarray}
for  the states in ${\bf 15}$.   

Although the decay channels are limited, they can be used to study many properties of beauty tetraquark 
$X_b$. We will show in the next section that with the above limited decay channels, one can extract a lot of 
information about the properties of $X_b$.

Several theoretical estimate of $X_b$ masses are higher than 5568 MeV. As pointed out in Ref.\cite{burns}  it is hard to 
understand that the $bsu$ baryons
$\Xi_b$ and $\Xi_b^*$ have masses of 5794 and 5945 MeV, but the $X_b$ states composed of $ \bar b s u \bar d$, $\bar b  d s \bar u$, and $\bar b u d \bar s$  
with an additional valence quark  would be hundreds of MeV lighter. 
Without a detailed dynamic model, this cannot be addressed. 
If it turns out that  
some of the states in $X_b$ have a mass larger than 5568 MeV, so that $X\to B K$ or even $X\to B \eta$ are also allowed kinematically. In this case, we have additional decay channels to study:
\begin{eqnarray}
&&A(X^{'}_{ds\bar u} \to B^0 K^-)=A(X^{'}_{ud\bar s} \to B^+ K^0)\nonumber\\
=&&-A(X^{'}_{su\bar d} \to B^+\bar K^0  ) = -A(X^{'}_{ud\bar s} \to B^0 K^+)= {1\over \sqrt{2}} \alpha^{'}\;,\nonumber\\
&&A(Y^{'}_{(d\bar d, s\bar s)u} \to B^0_s K^+) =A(Y^{'}_{(u\bar u, d\bar d)s} \to B^0 \bar K^0) \\
=&&
-A(Y^{'}_{(u\bar u, d\bar d)s} \to B^+ K^-) =-A(Y^{'}_{(u\bar u, s\bar s)d} \to B^0_s K^0) =  {1\over 2} \alpha^{'}\;,\nonumber\label{6k}
\end{eqnarray}
\begin{eqnarray}
&& A(Y^{'}_{(d\bar d, s\bar s)u} \to B^+\eta) = - A(Y^{'}_{(u\bar u, s\bar s)d} \to B^0\eta) =\sqrt{{3\over 8}} \alpha^{'}\;,
\label{6eta}
\end{eqnarray}
for the states in the $\bar 6$, and 
\begin{eqnarray}
&&A(X_{su\bar d} \to B^+ \bar K^0 ) = A(X_{ds\bar u} \to B^0 K^-) \nonumber\\
=&& 
A(X_{ud\bar s} \to B^+  K^0 ) = A(X_{ud\bar s} \to B^0 K^+)={1\over \sqrt{2}}\alpha\;,\nonumber\\
&& A(Y_{\pi s}\to B^+K^-)= -A(Y_{\pi s}\to B^0 \bar K^0) =  {1\over 2} \alpha\;,\nonumber\\
&&A(Y_{\eta d} \to B^0_s K^0) =  A(Y_{\eta s} \to B^0_s K^+) = -{1\over \sqrt{3}}\alpha\;,\nonumber\\
&&A(Y_{\eta s}\to B^0 \bar K^0) = A(Y_{\eta s} \to B^+ K^-) = {1\over 2\sqrt{3}}\alpha\;,\\
&&A(Z_{uu\bar s} \to B^+ K^+) =A(Z_{dd\bar s} \to B^0 K^0) \nonumber\\
=&& A(Z_{ss\bar d} \to B^0_s \bar K^0) =
 A(Z_{ss\bar u} \to B^0_s  K^-) =\alpha\;,\nonumber \label{15k}
 \end{eqnarray}
 \begin{eqnarray}
 &&A(Y_{\pi u} \to B^+ \eta) = - A(Y_{\pi d} \to B^0 \eta) = {1\over 2 \sqrt{3}}\alpha\;,\nonumber\\
&&A(Y_{\pi d} \to B^0 \eta) = - A(Y_{\pi u} \to B^+ \eta) = {1\over 2 \sqrt{6}}\alpha\;,\nonumber\\
&&A(Y_{\eta u} \to B^+ \eta) =  A(Y_{\eta d} \to B^0 \eta) = {1\over 6}(1+{5\over \sqrt{2}})\alpha\;,\\
&&A(Y_{\eta s} \to B^0_s \eta) =  {1\over 3 }(2+{1\over \sqrt{2}})\alpha\;,\nonumber \label{15eta}
\end{eqnarray}
for  the states in ${\bf 15}$.   

Note that in the above equations for decay amplitudes, there is a common overall factor $E_\pi/\sqrt{2} f_\pi$
from $v\cdot A$ term,  which we do not show explicitly in the above expressions.   
Also one needs to include the wavefunction  normalization factors, $\sqrt{M_X M_B}$ 
for the $X_b$ and $B$ fields as mentioned before.

From the leading order chiral Lagrangian with all the aforementioned factors correctly included,  one can use any one of the decay width, if measured, to determine the coupling constant $\alpha^\prime (\alpha)$.
For example we can evaluate the decay rate for $X_b\rightarrow B_s \pi^\pm$ as follows:
\begin{eqnarray}
{\cal M} ( X\rightarrow B_s \pi^\pm )  & = & \frac{\alpha'}{\sqrt{2} f_\pi} E_\pi 
\sqrt{M_X M_{B_s}}  \label{eq:amp}
\nonumber\\
\Gamma ( X\rightarrow B_s \pi^\pm ) &  = & \frac{\alpha^{'2}}{16 \pi} ~| \vec{p}_\pi |  ~
 \frac{M_{B_s}}{M_X} \left(  \frac{E_\pi}{f_\pi} \right)^2\; ,
\end{eqnarray}
We have used an approximation 
$E_\pi \approx (M_X - M_{B_s})$,  ignoring the recoil energy of $B_s$, and 
$p_\pi = \sqrt{E_\pi^2 - m_\pi^2}$.  Measurement of the decay width, the parameter $\alpha'$ can be determined.
Had the D0 data to be correct for the decay width to be 21.9 MeV  for this decay, one would obtain $| \alpha^{'} | \approx 1.3$ if $X_b$ belongs to ${\bf \bar 6}$.   Similarly $| \alpha | \approx 1.3$ if $X_b$ belongs
to the {\bf 15} representation.

\section{Discussions}

We now discuss how the $SU(3)$ properties of the beauty tetraquark states can be obtained by studying possible two-body decays with a $B$ meson and a light pesudoscalar octet meson in the final states.
If $SU(3)$ is at work in organizing the beauty tetraquark $X_b$ states, the $X_b$ states should come in the form of a complete 
irreduceable representation. This means that if a $X_b$ state composed of $\bar b s u \bar d$ or $\bar b d s \bar u$ as claimed by D0 is found, there should be other state come along in ${\bf \bar 6}$ or ${\bf 15}$. Then one should determined whether $X_b$ belongs to ${\bf \bar 6}$ or ${\bf 15}$. The obvious way is to find all possible states in a given
multiplet. To know whether a particular states are produced one has to look for its decay product the analyze the properties.
Therefore the decay channels are crucial in establishing the properties of the beauty tetraquark states. We now using the chiral Lagrangian constructed in section III to outline the strategy of carrying out the analysis.

From Eqs.(14) to (19), we see that if kinematically allowed, each of the beauty tetraquark state in ${\bf \bar 6}$ or ${\bf 15}$ can have two-body decays of the type $X_b \to B \Pi$. By analyzing the invariant mass of the final product one can identify whether there are resonant states which can be identified with a state in $X_b$.  The doubly charged tetraquark states $Z_{u u \bar d}$ and $Z_{u u \bar s}$ are the distinctive feature for that the beauty tetraquark to be a ${\bf 15}$ representation. 

It  D0 data is correct, only $X_b \to B \pi$ decays are kinematically allowed. In this case the available decay channels to analyze the $X_b$ state properties are limited. The allowed decay channels are given in eqs. (\ref{6pi}) and (\ref{15pi}).
We however note that a lot of information about $X_b$ can be extracted, in particular it is sufficient to determine whether the $X_b$ belongs to ${\bf \bar 6}$ or ${\bf 15}$.

Assuming $X_{s u\bar d, d s\bar u}$ have been discovered by analyzing $X_{s u\bar d}(X_{d s \bar u}) \to B^0_s \pi^+$ (or $B^0_s \pi^-$),
to determine whether $X_b$ belongs to ${\bf 15}$ or not is to find the doubly charged unique state $Z_{uu\bar d}$ which can decay into 
$B^+ \pi^+$ with a decay width twice as large as 
$X_{s u\bar d}(X_{d s \bar u}) \to B^0_s \pi^+$ (or $B^0_s \pi^-$).  Should this state exist, it can be detected. 
$Z_{dd\bar u} \to B^0 \pi^-$ can also add new information. 
Another type of decays is final states with $B^0 \pi^+$ from $Y_{\pi u}$ and $Y_{\eta u}$. This only occurs if $X_b$ belongs 
to the ${\bf 15}$ representation.  

Now we turn to final states which are allowed for both ${\bf \bar 6}$ and ${\bf 15}$ representations. 
We note that the ratio $R(B^+\pi^-/B^0_s \pi^0)$ of final states with $B^+ \pi^-$ and $B^0_s \pi^0$ 
can also provide useful information.

If $X_b$ belongs to ${\bf \bar 6}$, $Y^{'}_{(u \bar u, s\bar s) d} \to B^+ \pi^+$ and 
$Y^{'}_{(u\bar u, d\bar d)s} \to B^0_s \pi^0$ are relevant. The ratio $R(B^+\pi^-/B^0_s \pi^0) = 
\Gamma(Y^{'}_{(u \bar u, s\bar s) d} \to B^+ \pi^+)/\Gamma(Y^{'}_{(u\bar u, d\bar d)s} \to B^0_s \pi^0)$  is 1/2.
While for $X_b$ being in the ${\bf 15}$ representation, $Y_{\pi d} \to B^+ \pi^-$, $Y_{\eta d} \to B^+\pi^-$ 
and $Y_{\pi s} \to B^0_s \pi^0$ are relevant. If experiments can find the two separate states $Y_{\pi d}$ and 
$Y_{\eta d}$, this already tells that $X(5568)$ belongs to the ${\bf 15}$ representation. However, if experiments 
cannot distinguish them, the signal for $B^+\pi^-$ will be counted together. In this case the ratio 
$R(B^+\pi^-/B^0_s \pi^0)$ is 2/3 which is substantially different than the case with $X_b$ belongs to the  
${\bf \bar 6}$. 

Once the representation for $X_b$ has been determined, one can obtain all component  
state masses.   For those states which can decay into a pion in the final states, their masses 
can determined from experimental data. For those which do not have such decay channels, 
the sum rules derived from Eqs.(\ref{sum6}) and (\ref{sum15})  can be used to obtain 
the masses for those states.  The masses of those states that cannot be studied by $B\pi$
decays can be expressed in terms of the masses of those states that can be studied 
as the following for ${\bf \bar 6}$ or ${\bf 15}$,
\begin{eqnarray}
&&m(X^{'}_{ud\bar s}) = 2 m(Y^{'}_{(d\bar d,s\bar s)u }) - m(X^{'}_{ds\bar u})\;,\nonumber\\
&&m(Z_{ss\bar u})= m(Z_{ss\bar d}) =2 m(X_{su\bar d})- m(Y_{\pi \bar u})\;,\nonumber\\
&&m(Y_{\eta\bar s}) = m(Y_{\pi \bar s}) +m(Y_{\eta \bar u}) - m(Y_{\pi \bar u})\;,\\
&&m(X_{ud\bar s})=m(Z_{uu\bar s}) = m(Z_{dd\bar s}) = {1\over 2} \left (3m(Y_{\eta \bar s}) 
+ 4 m(Y_{\pi \bar u}) - 5 m(X_{su\bar d}) \right ) \;.\nonumber
\end{eqnarray}

We would also like to mention the possibility of using an off-shell light 
pseudoscalar which subsequent decay into two photons to test whether $X$ belongs to 
${\bf \bar 6}$ or ${\bf 15}$, taking $B_s^0\eta \to B_s^0 \gamma\gamma$ final state 
as an example. This can only happen if $X$ belongs to ${\bf 15}$ by $Y_{\eta s} \to B_s^0 \eta$ 
which is unlikely to be kinematically allowed. However, the $\eta$ has a large branching ratio 
(39\%)\cite{pdg} decay into $\gamma\gamma$, therefore $Y_{\eta s} \to B_s^0 \gamma\gamma$ 
may be possible to be observed. There are other states which can decay into $B_s^0\gamma\gamma$, such as $Y^{'}_{(u\bar u, d\bar d)s}, Y_{\pi s}\to B_s^0 \pi^0 \to B_s^0 \gamma\gamma$ with much larger decay widths. By removing events of the two $\gamma$ close to pion mass can help to single out events with the two $\gamma$s from $\eta$. This, however, also results in a much suppressed event number with $\Gamma(Y_{\eta s}\to B_s^0 \eta \to B_s \gamma\gamma)/ \Gamma(Y_{\pi s}\to B_s^0 \pi^0 \to B_s \gamma\gamma) \sim 4\times 10^{-5}$. Although $Y_{\eta s} \to B_s^0 \eta \to B_s^0 \gamma\gamma$ may be in principle provide useful information about the nature of the $X$, it is extremely difficult to carry out such tests experimentally.

If it turns out that  the $X_b \to BK$ and $X_b\to B\eta$ decay channels are also kinematically open, more information can be obtained.
For $X_b$ belongs to ${\bf \bar 6}$, with $X_b\to B\pi$ decay channels only, no direct information on $X^\prime_{ud\bar s}$ can be obtained. One needs to use mass relation to infer its mass. Now with $X_b\to B K$ channel opened, one can directly
measure this state by looking at the resonant structure of $B^+ K^0$ and $B^0K^+$ final states. The decay amplitudes for these two decay channels are
the same as $X^\prime_{su\bar d} \to B^0_s \pi^-$, but with smaller decay widths because of suppression of phase space.

For $X_b$ belongs to ${\bf 15}$,  if only $B \pi$ channels are kinematically allowed there are 
more states which cannot be studied directly probed by looking at decay products, 
for example the states, $X_{ud\bar s}$, $Y_{\pi s}$, $Y_{\eta s}$, $Z_{ss\bar u}$, $Z_{ss\bar d}$, 
$Z_{uu\bar s}$ and $Z_{dd\bar s}$ do not have $B \pi$ decay channels. We list possible decay channels these states can be studied once $X_b\to B K, \;\;B\eta$ decay channels are kinematically allowed in the following:
\begin{itemize}
\item When $X_b \to B K$ channels are open, $X_{ud\bar s} \to B^+ K^0, B^0K^+$ with  the 
same decay amplitude as $X_{su\bar d} \to B^0_s \pi^-$ can be used to study this state.

\item $Y_{\pi s}$ and $Y_{\eta s}$ can be studied by studying $B^+ K^-$ and $B^0 \bar K^0$ final states. If $Y_{\eta s} \to B^0_s \eta$ is also allowed, it can help probe $Y_{\eta s}$ properties.

\item The doubly charged state $Z_{uu\bar s}$ can be studied by $B^+ K^+$ final state.

\item The final states $B^0_s \bar K^0$ and $B^0_s K^-$ can provide information about 
$Z_{ss\bar d}$ and $Z_{uu\bar s}$ state, respectively.
\end{itemize}

\section{Conclusions}

In conclusion, we  studied properties of beauty tetraquark $X_b$ states with four different quarks and and 
their associated $SU(3)$ partner states. These states containing four different quarks should be in 
${\bf \bar 6}$ or ${\bf 15}$ of $SU(3)$ representations. 
We constructed the leading order chiral Lagrangian describing $X_b$ decays into a $B$ meson and a light 
pesudoscalar octet meson, and identified possible states which can be studied by these decays. 
Depending on the mass of $X_b$, the two-body decay channels $B K,\;B \eta$ may or may not be kinematically allowed. We find that evn just $B\pi$ decay channels are allowed, a lot of information about the properties 
$X_b$ can be extracted.   We discussed searching strategies to distinguish whether $X_b$ states belong to 
${\bf \bar 6}$ or ${\bf 15}$. We found that if $X_b$ belongs to ${\bf 15}$, there must be a new doubly charged 
four quark state which can decay into $B^+\pi^+$ with a decay width twice as large as that for 
$X^{(\prime)}_{su\bar d, ds \bar u}\to B_s^0 \pi^\pm$. 
There are several other features which can also distinguish whether $X_b$ belongs to ${\bf \bar 6}$ and 
${\bf 15}$.  We derived sum rules for $X_b$ mass useful in constructing all masses in a given representation. 
If the masses of $X_b$ allow $B K,\;B \eta$ decays to happen, more information about the $X_b$ states can 
be obtained.   As a final comment, we would like to point out that replacing the $\bar b$ quark by a $\bar c$ 
quark, one can obtain a similar class of states with the same $SU(3)$ properties\cite{partner}. 
Although the LHCb results did not confirm D0 finding of $X(5568)$ states, the interesting properties of $X_b$ 
are still worth investigating to understand more about hadron properties, we urge our experimental colleagues 
to carry out related searches.

\begin{acknowledgments}
X-G He was supported in part by MOE Academic Excellent Program (Grant No.~102R891505) 
and MOST of ROC (Grant No.~MOST104-2112-M-002-015-MY3), and in part by NSFC 
(Grant Nos.~11175115 and 11575111) and Shanghai Science and Technology Commission (Grant No.~11DZ2260700) of PRC. 
PK is supported by the National Research Foundation of Korea (NRF) Research Grant NRF-2015R1A2A1A05001869, and by SRC program of NRF Grant No. 20120001176 funded by MEST through Korea Neutrino Research Center (KNRC) at Seoul National University.
X.~G.~H. thanks Korea Institute for Advanced Study (KIAS) for their hospitality and partial support 
while this work was completed.
\end{acknowledgments}

\newpage

\end{document}